\documentclass[sigconf]{acmart}
\AtBeginDocument{%
  }

\copyrightyear{2026}
\acmYear{2026}
\setcopyright{cc}
\setcctype{by}
\acmDOI{10.1145/3815598.3815653}
\acmConference[FDG '26]{Foundations of Digital Games}{August 10--13, 2026}{Copenhagen, Denmark}
\acmBooktitle{Foundations of Digital Games (FDG '26), August 10--13, 2026, Copenhagen, Denmark}
\acmISBN{979-8-4007-2495-4/2026/08}

\usepackage{multirow}

\begin{document}

\title{LLMs are the Ideal Candidate for Mixed-Initiative Game Design Pillar Workflows}

\author{Julian Geheeb}
\email{julian.geheeb@tum.de}
\orcid{1234-5678-9012}
\affiliation{%
  \institution{Technical University of Munich}
  \country{Germany}
}

\author{Marvin Julian Schwarz}
\email{marvin.julian.schwarz@tum.de}
\affiliation{%
  \institution{Technical University of Munich}
  \country{Germany}
}

\author{Daniel Dyrda}
\email{daniel.dyrda@tum.de}
\affiliation{%
  \institution{Technical University of Munich}
  \country{Germany}
}

\author{Georg Groh}
\email{grohg@cit.tum.de}
\orcid{0000-0002-5942-2297}
\affiliation{%
  \institution{Technical University of Munich}
  \country{Germany}
}

\renewcommand{\shortauthors}{Geheeb et al.}

\begin{abstract}
  Game Design Pillars are natural language artifacts commonly used in game development to communicate a project's core vision and ensure a coherent player experience. 
  Their linguistic nature aligns well with the strengths of Large Language Models (LLMs), which excel at generating and interpreting natural language, making them strong candidates for supporting mixed-initiative workflows centered on design pillars. 
  In this study, we introduce a formal definition of game design pillars, present an initial prototype---SPINE---and investigate the utility of LLMs in the creation and decision-making processes associated with pillar-driven workflows. We begin with a pre-study to identify an appropriate model, comparing \texttt{gemini-2.0-flash} and \texttt{GPT-4o-mini}. Results show that Gemini is better suited to our tasks due to its greater output variety and consistency. 
  We then conduct a case study by deploying the tool at a local game jam. Findings indicate positive reception and clear value in integrating SPINE into early-stage development. Finally, we interview four experts, demonstrating the tool and allowing them to experiment with it in a controlled environment. While individual perspectives vary, the overall perception is encouraging and supports our intuition: LLMs can meaningfully contribute to game design pillar workflows. 
  These early findings highlight the potential of formalizing pillar-driven design as a research space and point toward several promising avenues for future work.
\end{abstract}



\received{16 December 2025}
\received[accepted]{01 February 2026}

\maketitle

\section{Introduction}
\textit{Lead with an example.}
That is the guiding principle behind this paper, and what we refer to as a \textit{design pillar}.
In traditional software engineering, design pillars such as \textit{Security} or \textit{Scalability} are used to guide system design toward higher quality~\cite{ahmad2025BasicPillarsSystemDesign, idunnu_paul2024FantasticFourSystemDesign}.
These pillars are often universal and broadly acknowledged, but they can also be tailored to a specific subset of systems, for example, Procedural Content Generation~\cite{lai2020towards}.

In contrast, game designers do not rely on universally agreed-upon pillars for every game.
Instead, pillars such as \textit{Combat} in \textit{God of War}~\cite{davis2018godofwar} or \textit{Realism} in \textit{Duskers}~\cite{keenan2017duskers} are defined separately to create unique and compelling experiences.
If pillars are instead the same between two or more games, players often refer to them merely as clones~\cite{despain2013100principles}.
Nevertheless, game design pillars remain a central tool for defining the guiding principles behind a game's vision, and subsequent decisions should adhere to these principles to form a cohesive experience~\cite{despain2013100principles}.
Working with pillars typically involves two key steps:
the \textit{creation of pillars}, resulting in natural language artifacts, and the \textit{usage of pillars}, which constitutes a decision-making process.
Both tasks align naturally with the strengths of LLMs---\textit{generating} and \textit{processing} natural language---making them an ideal candidate to support mixed-initiative pillar workflows, where \emph{ideal} denotes a promising fit worthy of evaluation rather than a proven outcome.

Yet, to our knowledge, no academic work investigates whether this relationship can yield positive improvements to design workflows.
In fact, the academic field largely refrains from discussing game design pillars at all: we found only two high-level sources~\cite{zagal2023considering,luo2021multidisciplinary}, despite pillars being well discussed and widely used in industry~\cite{Pears2017,davis2018godofwar,cleveland2019subnautica,keenan2017duskers}.
This disconnect limits knowledge transfer and slows the systematic analysis of game design practice.

To address this gap, our contributions are as follows:
\begin{itemize}
    \item To establish common terminology, we introduce a first \textit{formal} definition of game design pillars (\autoref{sec:pillars}) and provide a dataset of 55+ real-world examples (\autoref{sec:pillar_set}).
    \item We build a prototype, SPINE, with a basic user interface (\autoref{sec:spine}) to support both the \textit{creation} and \textit{decision-making} phases of pillar workflows using LLMs.
    \item We evaluate SPINE using three approaches: a performance evaluation of pillar creation using curated datasets (\autoref{sec:model_performance}), a demonstration via a small-scale case study (\autoref{sec:selfstudy}), and a qualitative study with two studios (\autoref{sec:interviews}).
\end{itemize}

\section{Game Design Pillars}\label{sec:pillars}
During our research, we compiled a small dataset of documented design pillars from existing games, presented in \autoref{sec:pillar_set}.
As previously mentioned, game design pillars serve as foundational principles that guide decision-making throughout development.
Design pillars can be understood as self-imposed constraints on the design space intended to support a coherent and focused player experience.
New features, mechanics, or assets are typically evaluated against these pillars to ensure alignment with the intended vision.
Design pillars are commonly established at the beginning of a project~\cite{despain2013100principles}, as modifying them during development may invalidate prior decisions and undermine coherence.
They also play an important communicative role, helping teams maintain a shared understanding of design goals---especially critical in large-scale development teams~\cite{zagal2023considering}.
Concurrent findings suggest recurring challenges in pillar workflows, including vagueness, conflicting interpretations, difficulties applying pillars to design decisions, and insufficient documentation structure~\cite{Dyrda2026GameDesignPillars}.
In practice, pillars often articulate target emotions and player experiences (e.g.,~\cite{kara2021GamePillars,keenan2017duskers}) or define core mechanics and structural design commitments (e.g.,~\cite{cleveland2019subnautica,davis2018godofwar}).

To establish a shared vocabulary, we introduce a formal definition of game design pillars.
This formalization enables their use in a computational context, as pillars must be expressed in a structured manner before they can be systematically generated, analyzed, or processed.
Unless stated otherwise, all subsequent references to design pillars in this work are grounded in this definition.
The proposed definition is derived from a comparative review and synthesis of design pillars described in both academic and industry sources, including
\cite{despain2013100principles,cleveland2019subnautica,davis2018godofwar,keenan2017duskers,graft2012diablo,lapikas2012deusex,ali2013destiny,cain2023fallout,ParadoxInteractive_GamePillars,kara2021GamePillars,Wagar_DesignPillars_GameDesignSkills,Pears2017,zagal2023considering}.

\begin{definition}[Game Design Pillar]\label{def:pillars}
    A \emph{game design pillar} is a normative design construct that functions as a high-level principle for directing and constraining decision-making in game development.
    It is formally composed of:
    \begin{enumerate}
        \item a succinct title that names the principle,
        \item an expository statement that specifies the intended experiential or structural property the game should embody.
    \end{enumerate}
\end{definition}
\noindent
In this context, a \emph{structural property} refers to a characteristic of the game’s formal system, including its mechanics, dynamics, progression architecture, interaction loops, or rule-based organization. 
By contrast, an \emph{experiential property} refers to the intended emotional, cognitive, or aesthetic experience of the player---in short, the player experience.

\paragraph{Quality Criteria}\label{sec:quality_criteria}
The following are qualitative properties of well-formed pillars:
\begin{itemize}
    \item \textbf{Clarity:} The meaning of the principle should be unambiguous to stakeholders.
    \item \textbf{Unicity:} Each pillar should articulate only one coherent design principle.
    \item \textbf{Conciseness:} The pillar should be expressed in minimal, economical language.
    \item \textbf{Actionability:} The principle should be applicable to concrete design decisions.
\end{itemize}

\paragraph{System Constraints}\label{sec:system_constraints}
A set of game design pillars should satisfy the following system-level constraints in order to remain analytically useful, internally coherent, and practically applicable:
\begin{itemize}
    \item \textbf{Mutual non-contradiction:} Individual pillars must not negate or undermine one another.
    \item \textbf{Completeness:} The set should collectively articulate the game’s core experiential goals without leaving critical aspects of the intended experience unspecified.
    \item \textbf{Bounded size:} The number of pillars should remain deliberately limited and small. 
\end{itemize}
A typical set size contains three to five pillars, depending on the scope of the project.
More fine-grained considerations may be addressed through subordinate or derivative sets.

\section{SPINE}\label{sec:spine}
\begin{figure*}
  \centering
  \includegraphics[width=\textwidth]{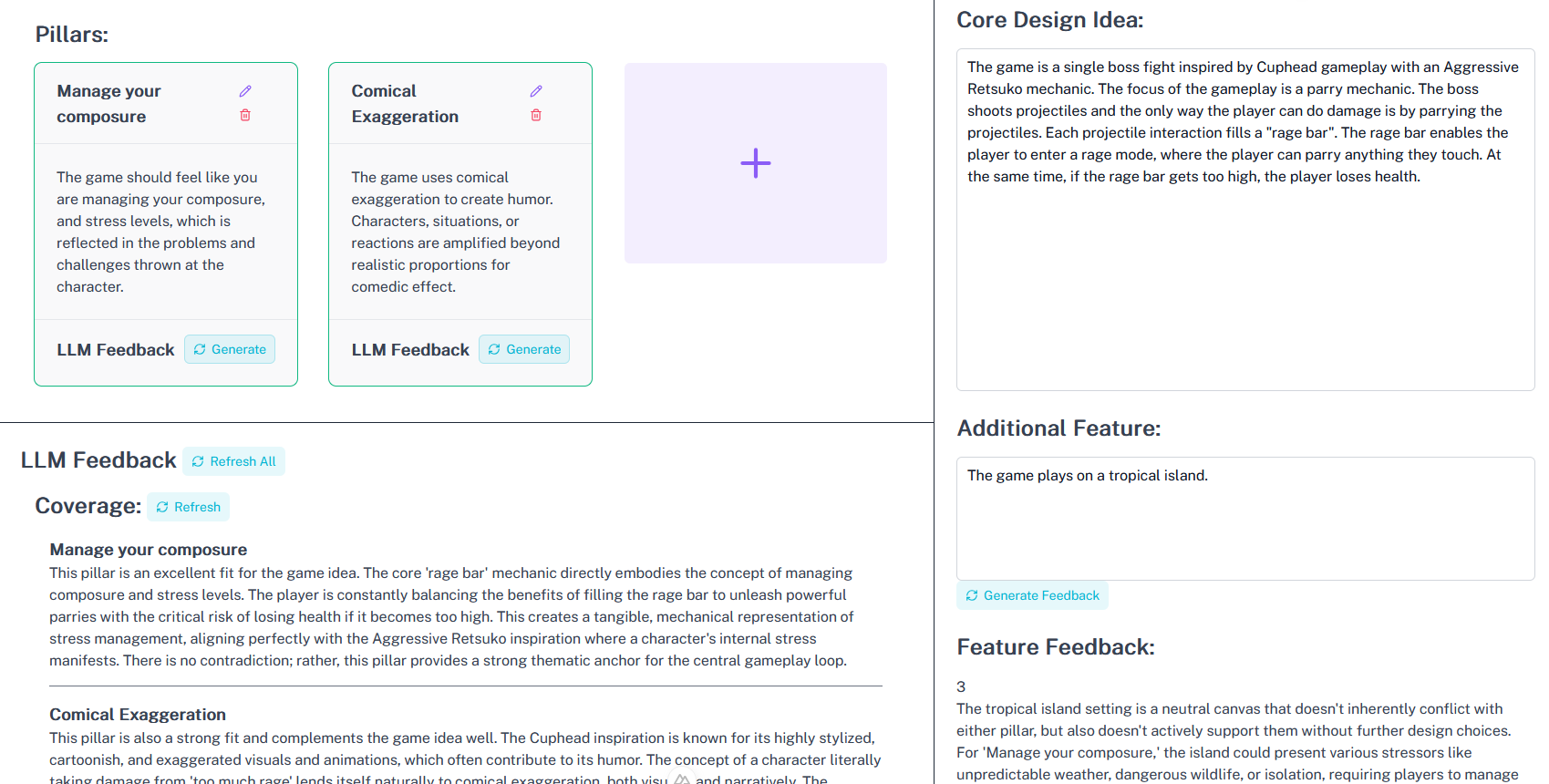}
  \caption{A screenshot of SPINE's user interface.}
  \Description{A graphic showing how game design pillars are used in the design process.}
  \label{fig:spine}
\end{figure*}

Based on our findings presented in \autoref{sec:pillars} and \autoref{def:pillars}, we developed our first proof-of-concept, \textit{SPINE}, a mixed-initiative \textbf{S}ystem for \textbf{P}illar-based \textbf{IN}teractive \textbf{E}xperience design.
The prototype’s architecture is divided into a \textit{Django}\footnote{https://www.djangoproject.com/} backend, chosen for its robustness, and a \emph{Nuxt4}\footnote{https://nuxt.com/} frontend, chosen for its accessibility and its use of \emph{NuxtUI}\footnote{https://ui.nuxt.com/} components.
The backend communicates with an LLM through its respective API, which can technically be swapped out for any API-based LLM to support different experiments.
The frontend provides a minimal user interface (see \autoref{fig:spine}), where users can create three different types of content:
\begin{itemize}
    \item A core design idea, containing a high-level description of the game.
    \item A set of pillars, each consisting of a title and a description.
    \item A feature idea, containing a description of the feature.
\end{itemize}
Additionally, SPINE integrates several LLM-based functionalities to address common challenges when working with design pillars.
For this first prototype, we focused on the \textit{pillar creation} step of the workflow, but also support the \textit{decision making} step at a basic level.
All prompts can be found in \autoref{sec:prompts}.
In the following, we present each LLM-powered feature.

\subsection{Structural Pillar Analysis}
The structural analysis provides feedback to the user on whether an individual pillar is well formed.
As a first approach, we created a single prompt that checks for structural issues based on \autoref{def:pillars} (Title, Format) and two quality criteria from \autoref{sec:quality_criteria} (Clarity, Focus):
\begin{itemize}
    \item Does the title match the description? (Title)
    \item Is the description written as continuous text? (Format)
    \item Is the intent of the pillar clear? (Clarity)
    \item Is the pillar focused on one aspect? (Focus)
\end{itemize}
We use continuous text as a practical lower bound for an expository statement without imposing unnecessary stylistic constraints.
For this initial research, we omitted the criteria \emph{Conciseness} and \emph{Actionability}, as they usually require additional domain-grounded interpretation.
After prompting the system, the LLM provides structural feedback on the four issues, indicating whether an issue is present and rating its severity on a scale from 1 (low) to 5 (high).

\subsection{Structural Pillar Repair}
If, during the structural analysis, the system identifies structural issues in the current pillar, users can opt to address these issues with the help of the LLM.
The user can subsequently decide whether to keep the original version or replace the pillar with the LLM-improved version.
In any case, the user can always edit the pillar, regardless of whether it is their own or the LLM-generated version.

\subsection{Pillar Set Validation}
This feature set focuses on the system constraints (see \autoref{sec:system_constraints}), specifically mutual non-contradiction and completeness.
We leave out limited set size, as this can be implemented programmatically without the use of LLMs.
We created three distinct prompts, each combining the core design idea with the pillars and focusing on one specific issue:
\begin{itemize}
    \item \textbf{Coverage:} The LLM is asked to evaluate whether the given pillars are a good fit for the core design idea. This prompt checks for incomplete coverage of the core design idea.
    \item \textbf{Contradictions:} The LLM is asked to weigh all pillars against each other to check for possible contradictions. This prompt checks for mutual non-contradiction.
    \item \textbf{Additions:} The LLM is asked to suggest additional pillars for the core design idea that are currently missing. This prompt checks for completeness.
\end{itemize}
Each response is provided in plain text to allow for reasoning and includes a short explanation of why the model identifies a potential issue.

\subsection{Feature Validation}
Finally, this feature provides a conceptual entry point into the \textit{decision making} step of working with pillars.
The LLM is prompted to evaluate the feature idea against the given pillars.
The feature is rated on a scale from 1 (low) to 5 (high) based on how well it fits the pillars, and the model provides an explanation of its decision.

\section{Evaluation and Results}\label{sec:results}
We evaluated SPINE and its overall potential by adopting a broad evaluation approach that covers multiple aspects of the system.
Accordingly, we employed three evaluation methods corresponding to Types~1--3 of Ledo et al.'s~\cite{ledo2018evaluation} evaluation strategies for HCI toolkits, with each method addressing a different aspect of the tool.
The model comparison (\autoref{sec:model_performance}) should be understood as a pre-study rather than a full-scale evaluation of SPINE’s capabilities.
The order in which the evaluations are presented reflects the sequence in which the studies were conducted.

\subsection{Exploratory Model Comparison}\label{sec:model_performance}
This pre-study was primarily conducted to select the first LLM candidate for our later evaluations, but it also allowed us to gain an overview of LLMs' general understanding of game design.
We compared the performance of two models: \texttt{gemini-2.0-flash}\footnote{https://docs.cloud.google.com/vertex-ai/generative-ai/docs/models/gemini/2-0-flash} (Gemini) and \texttt{GPT-4o mini}\footnote{https://platform.openai.com/docs/models/gpt-4o-mini} (GPT).
The models were chosen to achieve fast runtime and high cost efficiency, as the features of SPINE are intended to be used in an iterative process and assumed to be used frequently.
As there is currently no existing dataset, baseline, or benchmark, our approach is exploratory by design.
Future work includes the creation of a proper dataset, which would allow for comparison between iterations and approaches.

\subsubsection{Methodology}
We started by creating two sets of pillars, each consisting of three pillars.
The first set originates from a student project called \emph{Ordinary}, which was developed in parallel with SPINE.
The second set is derived from the popular game \emph{Sea of Thieves}\footnote{https://www.seaofthieves.com/} by reverse engineering the game's main features.
We selected \emph{Sea of Thieves} instead of a game from our small dataset in \autoref{sec:pillar_set} because we needed a game with which we were familiar in order to express informed qualitative thoughts about the models' output.
After creating the pillar sets, we ran experiments for each functionality of SPINE.

\paragraph{Structural Pillar Analysis and Repair}
The goal was to evaluate the robustness of the structural pillar analysis for each model.
Therefore, we first prompted each model three times for each pillar to determine whether the model would provide consistent feedback or whether it would ignore the structure entirely and effectively produce random feedback.
Afterward, for each model, we generated an improved version of each pillar once and again prompted the model three times to provide a structural analysis of its own improved pillar.
The results reveal the model’s consistency and whether it converges on its own revised pillar or continues iterating.
We recorded the scores returned by the model for each pillar and issue (1--5).

\paragraph{Pillar Set and Feature Validation}
For each remaining feature, we prompted the model once for each game idea.
For the pillar set validation, we provided the system with a core design idea and analyzed its output qualitatively.
For the feature validation, we entered a new feature idea for \emph{Ordinary} and an already existing feature for \emph{Sea of Thieves}.

\subsubsection{Results and Discussion}
For this section, we combine the presentation of the results with a short discussion to keep the pre-study short and concise. 

\paragraph{Structural Pillar Repair}
An immediate observation was that GPT consistently produced longer responses, often expanding descriptions into explanatory texts rather than concise pillar statements.
Gemini, by contrast, reformulated the pillars more succinctly.
Furthermore, both models converged on transforming the first and last pillars of \emph{Sea of Thieves} into variations of player agency, thereby reducing the distinctiveness of the set and obscuring the specific design focus of the game.
GPT generally kept the original titles and elaborated on the descriptions with additional explanations.
Gemini, by contrast, preferred reformulations that mapped the pillars to more standardized terminology.

\paragraph{Structural Pillar Analysis}
The resulting ratings for all prompts are displayed in \autoref{tab:gemini} for Gemini and \autoref{tab:gpt} for GPT.
Both models demonstrated distinct tendencies in how they handled pillar analysis and improvement. 
GPT would generally return the same structural issues regardless of the pillar or whether the pillar had been fixed by the model itself.
This pattern suggests potential limitations in GPT’s ability to fully interpret the given task. 
Gemini remained largely consistent across the evaluation process, showed variation between pillars, and demonstrated a notable improvement in its own scores when assessing the revised pillars, removing nearly all warnings in the process. 

\begin{table}[t!]
    \centering
    \caption{Pillar Evaluation Summary for Gemini. Each cell presents three different issue severity ratings for the same pillar. '-' indicates no structural issues found.}
    \label{tab:gemini}
    \resizebox{\columnwidth}{!}{
    \begin{tabular}{p{0.3\columnwidth}ccccc}
        \hline
        \textbf{Pillar} & \textbf{Version} & \textbf{Title} & \textbf{Clarity} & \textbf{Focus} & \textbf{Format} \\
        \hline

        \multirow{2}{*}{
            \begin{minipage}{0.3\columnwidth}
                Choose your Journey
            \end{minipage}} 
            & Original & - | - | - & 3 | 3 | 3 & 2 | 2 | 2 & - | - | - \\
            & Improved & - | - | - & - | - | - & - | - | - & - | - | - \\
        \hline

        \multirow{2}{*}{
            \begin{minipage}{0.3\columnwidth}
                Kintsugi Storytelling
            \end{minipage}} 
            & Original & 3 | 3 | 4 & 4 | 2 | 3 & - | - | 2 & - | - | - \\
            & Improved & - | - | - & - | - | - & - | - | - & - | - | - \\
        \hline
        
        \multirow{2}{*}{
            \begin{minipage}{0.3\columnwidth}
                Moments that Matter
            \end{minipage}} 
            & Original & 3 | 3 | 3 & 4 | 4 | 4 & 3 | 2 | 2 & - | - | - \\
            & Improved & - | - | - & - | - | - & 3 | 3 | 3 & - | - | - \\
        \hline
        \hline
        
        \multirow{2}{*}{
            \begin{minipage}{0.3\columnwidth}
                Choose your own Adventure
            \end{minipage}} 
            & Original & - | - | - & - | 3 | - & 3 | 2 | 3 & - | - | - \\
            & Improved & - | - | - & - | - | - & - | - | - & - | - | - \\
        \hline
        
        \multirow{2}{*}{
            \begin{minipage}{0.3\columnwidth}
                Dynamic Open World
            \end{minipage}} 
            & Original & - | - | - & - | - | - & 3 | 3 | 3 & - | - | - \\
            & Improved & - | - | - & - | - | - & 3 | 3 | 3 & - | - | - \\
        \hline
        
        \multirow{2}{*}{
            \begin{minipage}{0.3\columnwidth}
                Freedom of Conduct
            \end{minipage}} 
            & Original & - | - | - & 3 | 3 | 3 & 2 | 2 | 2 & - | - | - \\
            & Improved & - | - | - & - | - | - & - | - | - & - | - | - \\
        \hline

    \end{tabular}
    }
\end{table}

\begin{table}[t!]
    \centering
    \caption{Pillar Evaluation Summary for GPT. Each cell presents three different issue severity ratings for the same pillar.}
    \label{tab:gpt}
    \resizebox{\columnwidth}{!}{
    \begin{tabular}{p{0.3\columnwidth}ccccc}
        \hline
        \textbf{Pillar} & \textbf{Version} & \textbf{Title} & \textbf{Clarity} & \textbf{Focus} & \textbf{Format} \\
        \hline

        \multirow{2}{*}{
            \begin{minipage}{0.3\columnwidth}
                Choose your Journey
            \end{minipage}} 
            & Original & 3 | 3 | 3 & 4 | 4 | 4 & 3 | 3 | 3 & 2 | 2 | 2 \\
            & Improved & 3 | 3 | 3 & 2 | 4 | 4 & 3 | 2 | 3 & 1 | 2 | 1 \\
        \hline

        \multirow{2}{*}{
            \begin{minipage}{0.3\columnwidth}
                Kintsugi Storytelling
            \end{minipage}} 
            & Original & 3 | 3 | 3 & 4 | 4 | 4 & 3 | 3 | 3 & 2 | 2 | 2 \\
            & Improved & 3 | 3 | 3 & 3 | 4 | 3 & 2 | 4 | 3 & 2 | 2 | 2 \\
        \hline
        
        \multirow{2}{*}{
            \begin{minipage}{0.3\columnwidth}
                Moments that Matter
            \end{minipage}} 
            & Original & 3 | 3 | 3 & 4 | 4 | 4 & 3 | 3 | 3 & 2 | 2 | 2 \\
            & Improved & 3 | 3 | 3 & 4 | 4 | 4 & 3 | 3 | 4 & 2 | 2 | 2 \\
        \hline
        \hline
        
        \multirow{2}{*}{
            \begin{minipage}{0.3\columnwidth}
                Choose your own Adventure
            \end{minipage}} 
            & Original & 3 | 3 | 3 & 4 | 4 | 4 & 4 | 3 | 3 & 2 | 2 | 2 \\
            & Improved & 3 | 3 | 3 & 2 | 2 | 2 & 3 | 3 | 3 & 2 | 2 | 2 \\
        \hline
        
        \multirow{2}{*}{
            \begin{minipage}{0.3\columnwidth}
                Dynamic Open World
            \end{minipage}} 
            & Original & 3 | 3 | 3 & 4 | 4 | 4 & 4 | 3 | 4 & 2 | 2 | 2 \\
            & Improved & 3 | 3 | 2 & 3 | 4 | 3 & 4 | 4 | 3 & 2 | 3 | 2 \\
        \hline
        
        \multirow{2}{*}{
            \begin{minipage}{0.3\columnwidth}
                Freedom of Conduct
            \end{minipage}} 
            & Original & 3 | 3 | 3 & 4 | 4 | 4 & 4 | 3 | 4 & 2 | 2 | 2 \\
            & Improved & 3 | 3 | 3 & 4 | 4 | 4 & 3 | 4 | 4 & 2 | 2 | 2 \\
        \hline

    \end{tabular}
    }
\end{table}

\paragraph{Pillar Set and Feature Validation}
For coverage, the output aligns closely with expectations: in both cases, the game idea is well represented by the existing pillars, and each pillar clearly supports the intended design direction.
Contradictions revealed the biggest difference between the models, with Gemini identifying multiple reasonable conflicts, while GPT did not report any in the case of \emph{Ordinary}.
Furthermore, for \emph{Sea of Thieves}, neither model identified potential contradictions caused by two pillars being overly similar.
This suggests a limitation in the prompt design: the models were guided to search for explicit contradictions but not for duplication or redundancy within the pillar set.
For the additions, GPT proposed a greater number of additions than Gemini in both cases.
All additions from both models were reasonable.
For the feature validation, there was no major difference between the models.
Both GPT and Gemini were able to clearly link the ideas to the existing pillars for all our inputs.

\paragraph{Short Discussion}\label{sec:short_discussion}
Overall, we did not observe any major differences between the models in terms of the Coverage, Contradictions, Additions, and Additional Feature functionalities.
However, Gemini outperformed GPT in the structural analysis in terms of issue detection and issue fixing, as GPT would effectively give all pillars the same rating regardless of whether they had been fixed.
Therefore, we decided to proceed with Gemini for all subsequent evaluations.
While the results clearly indicate room for improvement in terms of output quality, the goal of the pre-study was not yet to find the best way to utilize LLMs. Rather, we set out to identify a suitable candidate for a first prototype, which we successfully achieved.

\subsection{Small-scale Case Study}\label{sec:selfstudy}
Given the tool’s early stage of development, we conducted a small-scale exploratory case study with self-study elements, followed by short participant interviews.
One researcher participated in a local 42h game jam, using SPINE as the primary design tool to develop a game in collaboration with two student participants.
While development was collaborative, only the researcher directly interacted with and managed the SPINE tool, while the students contributed through discussion, ideation, and feedback.
The researcher’s role was to facilitate SPINE’s use and document the process.
After the game jam, we conducted brief interviews with the student participants individually to capture their perspectives.
The goal of this approach was to illustrate SPINE’s workflows and explore its effectiveness in supporting design pillar creation and decision-making in a realistic, time-constrained context.
This study does not seek to assess user learning or adoption, but rather to provide formative insights into the toolkit’s functionality, expressiveness, and design trade-offs.
We acknowledge potential biases arising from the researcher’s dual role as tool designer and primary user. 
However, combining self-study with participant interviews aligns with established exploratory evaluation practices in early-stage HCI tool research.

\subsubsection{Participants}
Two participants (P1--P2) took part in the study.
P1 was a 26-year-old female working student at an indie studio and a master's student in games engineering.
P2 was a 23-year-old male working student at a different indie studio and a master's student in games engineering.
Both participants reported prior familiarity with game design pillars and experience using them in practice.

\subsubsection{Approach}\label{sec:approach}
Before the game jam, we created a concrete list of questions that we aimed to address:
\begin{itemize}
    \item How does the tool support the creation of design pillars in the early stages of development?
    \item How does the tool support decision-making during later stages of development?
    \item What are the limitations?
    \item Which functionality is missing?
\end{itemize}
This study follows a research-by-design approach, as the goal was to inform future iterations of the toolkit.

As game jams are typically fast paced and time constrained, we did not define a strict step-by-step action plan.
Instead, we formulated a set of guiding rules intended to help answer the questions:
\begin{itemize}
    \item Use the tool whenever the discussion lends itself to it.
    \item Document all steps of the initial design phase.
    \item Update the documentation every 1--2 hours.
    \item Include the team’s responses to the generated feedback.
\end{itemize}

The first rule was intended to avoid collecting insufficient data due to a lack of familiarity with the tool.
Even when a decision had already been made within the team, we still used the tool to observe its feedback.
Additionally, instances in which the tool felt inadequate were treated as informative, as they helped identify specific shortcomings and guided future iterations.

\subsubsection{Results}
The results are divided into the researcher’s documentation of the process and post-jam interviews with the two student participants. 

\paragraph{Usage Documentation}
The game jam began on Friday at 18:00 with the announcement of the theme: the meme \textit{This Is Fine}, depicting a character asserting calm despite an unfolding disaster.\footnote{https://knowyourmeme.com/memes/this-is-fine}

In the evening, the team began by creating an initial design pillar directly from their interpretation of the theme and requesting SPINE’s AI feedback.
The feedback was largely accepted, but the team attempted manual refinement before using SPINE’s pillar-improvement function, which resulted in the pillar \emph{embrace sarcastic resilience}.
Next, a preliminary mechanic idea (a ``stress meter'') was entered into the core design idea field, prompting SPINE to suggest an additional pillar, \emph{unleash controlled rage}.
When SPINE flagged a contradiction between the two pillars, the team initially disagreed but continued iterating.
After further refinement, the core design idea was finalized and expanded, and SPINE was used to suggest and validate additional pillars.
This resulted in four pillars, including variations on rage, composure, and risk--reward dynamics.
SPINE again identified contradictions, which led the team to reconsider the terminology and scope of the pillars.
Renaming ``rage'' to ``composure'' reduced conflict but highlighted that the overall pillar set was too broad for the project’s scale.
Consequently, the team discarded the previous pillars and rebuilt them from scratch.
They authored a new pillar, \emph{manage your composure}, which SPINE attempted to repair but did not improve to the team’s satisfaction.
For the second pillar, an initial \emph{sarcastic humor} concept was refined with AI assistance into \emph{comical exaggeration}.
A final SPINE check confirmed full coverage and no contradictions, after which the team moved to prototyping.

Early in development, the team explored a workplace setting involving constant interruptions from emails, calls, and a hostile boss.
Instead of formalizing this as a pillar, the setting was described using SPINE’s feature evaluation text field, which returned a 5/5 compatibility rating with a detailed justification.
Alternative settings and varying levels of description granularity were tested, producing lower scores that aligned with the team’s expectations.
The following day, during prototype development, the team debated whether the game space should be vertically expansive or tightly constrained.
Both positions were entered into SPINE’s feature feedback system to assess alignment with the established pillars.
SPINE favored the constrained layout, citing stronger support for composure management.
While participants questioned parts of the explanation, the interaction led to the suggestion that future AI feedback should explicitly reference the evaluated text to improve interpretability.
After this discussion, SPINE was no longer used, as the team had reached a stable design vision and no further high-level decisions were required before the submission deadline.

\paragraph{Participant Reflection}
P1 expressed an overall positive view of the tool, describing it as practical and supportive of the design process: “Generally, the tool reassures you about your ideas and raises questions, but it also gave us ideas in the case of pillar additions.”
Similarly, P2 noted that the tool was particularly helpful for defining design pillars.
Both participants agreed that SPINE supports structuring design thinking and articulating ideas more clearly.

At the same time, both participants observed that the LLM-generated rephrasings were occasionally overly generalized or verbose.
P1 emphasized that this was not a major issue, as pillars could be edited after accepting AI-generated suggestions.
P2 suggested that adjusting the complexity of the output could improve clarity and ease of understanding.
Both participants also expressed a desire for additional guidance on formulating design pillars, such as short explanations, examples, or hints.
They noted that such guidance could benefit both novice and experienced developers by making the system’s internal interpretation of pillars more transparent.
Relatedly, P1 suggested that the tool could actively guide users through pillar creation by posing reflective questions or restating its interpretation of user input to enable more precise rephrasing.

Feedback on the additional feature evaluation functionality was more mixed.
P1 reported that it worked well and expressed interest in extensions such as collecting and comparing multiple features or allowing the LLM to generate feature ideas, similar to the pillar suggestion mechanism.
P2 noted that the feature evaluation was used less frequently, but still found the feedback useful in the instances where it was applied.
He attributed the limited usage partly to the time pressure of the game jam, observing that checking the alignment of new ideas with existing pillars often happens implicitly.
However, P2 emphasized that in larger teams, where not all members are deeply involved in design decisions, such functionality could serve as an effective preliminary filter, enabling quick validation of ideas before broader team discussion.

\subsection{Expert Interviews}\label{sec:interviews}
To gain insight into how the tool is perceived by practitioners, we conducted semi-structured qualitative interviews with four game developers from two different studios.
All interviews were conducted in person, audio recorded, and subsequently transcribed.
In addition, we took observational notes during the sessions, capturing participants’ comments and reactions.
The interviews were conducted in English, although the participants’ native language was German, and each session lasted approximately 50--60 minutes.
Prior to the interviews, we emphasized the importance of honest and direct feedback in order to better understand both the strengths and shortcomings of SPINE.
Participants then read and signed a consent form, after which demographic information was collected via a questionnaire.

\paragraph{Participants}
Four participants (P3--P6) took part in the interview study.
P3 was a 24-year-old male game designer with one year of professional experience, and P4 was a 32-year-old male game developer with five years of experience.
P3 and P4 worked at the same game development studio, which has been active for over five years and was, at the time of the study, developing its second commercial title.

P5 was a 23-year-old female with a background in games engineering and two years of experience in game art freelancing.
P6 was a 27-year-old male co-founder and game designer with approximately 1.5 years of experience in game development and design.
P5 and P6 were collaborating on their first commercial title under P6’s studio.

\subsubsection{Interview Process}
In the first phase of the interview, we asked participants general questions about design pillars, including whether and how they use them in practice.
This initial discussion was intended to establish a shared understanding of design pillars and to elicit participants’ perspectives before introducing the tool, thereby reducing potential bias.

In the second phase, we demonstrated the full functionality of SPINE using a laptop brought to the interview.
We adopted a walkthrough-style demonstration to emphasize the tool’s utility rather than usability, as SPINE is an early-stage artifact.
The demonstration began with a predefined set of pillars from \autoref{sec:selfstudy}, followed by the creation of a new pillar to illustrate the pillar authoring workflow and structural analysis.
We then demonstrated the design and feature evaluation functionality.
Throughout the demonstration, we addressed participants’ questions and posed occasional follow-up questions.
This phase concluded with a request for participants’ initial impressions of the tool.

Next, participants were given the opportunity to explore SPINE hands on by creating a new project and defining their own pillars and game ideas.
This phase lasted approximately 30 minutes.
Given that formulating design pillars is a complex task unlikely to be completed within a short timeframe, the goal was not completion but familiarization with the workflow, receiving feedback on their own ideas, and assessing whether this interaction altered their initial impressions.

Finally, we revisited participants’ impressions after their hands-on experience and concluded each interview with an open-ended discussion.
This discussion focused on the perceived potential of the tool, its applicability to participants’ professional practice, the strengths and weaknesses of specific features, and any missing functionality they would have expected.

\subsubsection{Results}
All participants reported prior familiarity with design pillars before the interview.
Their descriptions broadly aligned with common definitions of design pillars, although several participants expressed uncertainty in articulating them precisely.
For example, P4 described design pillars as ``[t]he core foundations that your game is built upon, like the core mechanics. Doesn't have to be the mechanics though, it could also be the core vibes, core principles that your experience is built upon''.
Furthermore, P5's definition was more aligned with aspects of a game concept~\cite{geheeb2025diamonds} than with the common pillar definition presented in \autoref{sec:pillars}.
All participants had previously worked with design pillars at least once.
P3 reported using them frequently as a tool for communicating a project’s vision.
P4 described applying design pillars during the reworking of a previous game.
P5 and P6 had each used design pillars once; P6 noted that their relevance decreased over the course of the project.
Despite differences in experience and frequency of use, all participants described design pillars as a useful tool within their design practice.

At the start of Phase~2, P4 stated that he was ``kind of biased against LLMs'' due to ethical concerns.
When asked to elaborate, he mentioned issues such as the replacement of jobs, distrust toward the companies controlling LLMs, and their environmental impact.
During the demonstration phase, P3, P5, and P6 primarily listened and asked clarifying questions.
In contrast, P4 actively commented on the LLM-generated text for each feature.
He criticized the feedback, stating that it largely rephrased the input without adding substantive suggestions.
At other times, however, he acknowledged agreement with the output, describing some feedback as ``fair.''

When asked about their first impressions, P3 responded positively, stating ``I think that's very cool, I actually like that.''
He further noted that he appreciated being able to establish a core design idea independently before developing design pillars based on it.
P5 expressed a similar view, emphasizing the value of writing ideas down and describing the tool as helpful due to its cross-referencing with the game concept.
P4 also responded positively to the structure connecting core design ideas and design pillars, while reiterating that, in his view, most weaknesses stemmed from the use of LLMs rather than from the overall tool.
He added that in the absence of a team to brainstorm with, the tool could still provide an initial perspective.
P6 stated that the feedback generally made sense, although he expressed uncertainty regarding a newly generated pillar, noting that it appeared to involve only minor wording changes.

\paragraph{P3 Observations \& Second Impressions}
P3 began the session by describing his core design idea, followed by the formulation of an initial design pillar.
After receiving feedback on the first pillar, he reflected on the generated revision:

\begin{quote}
``I kept this pillar vague on purpose with vibe. And I can see the generated version is more specific how this would be achieved, which is more helpful so I take this one.''
\end{quote}

For the second pillar, P3 noted that the feedback introduced new elements beyond his original intent.
While he did not fully adopt the suggestion, he reported appreciating its phrasing and chose to edit the AI-generated version rather than discard it entirely.

When interacting with the design evaluation functionality, P3 expressed a preference for feedback that highlights omissions rather than proposing entirely new pillars:

\begin{quote}
``I would prefer it if it would poke me in the direction of `hey this is missing' instead of completely suggesting a new pillar. I would prefer that for my workflow, and it would be more respectful towards me. But it's good that it suggests me an aesthetics pillar, [...]''
\end{quote}

Following this suggestion, P3 added an aesthetics pillar, describing an intentional twist that juxtaposed nature and technology.
While revising this pillar and receiving formatting feedback, he explained how he preferred to integrate system input into his process:

\begin{quote}
``So I think what I would do usually is to take these feedback points, which I think are valid, and try to fix them myself first, in order to prevent [...] too many parts of the creative process [...] to be done by the system [...] and instead use the system like to poke at me and say `ok you are not working as cleanly as you should', which I think is great.''
\end{quote}

When evaluating feedback on a feature that incorporated this twist, P3 disagreed with the system’s response:

\begin{quote}
``So in this case it was a deliberate attempt by me to have a very unique sort of twist in it. And I do feel like it is just saying `it shouldn't have a twist', instead of actually evaluating the twist. So I don't agree at all with that, which is fine, but I would have wished for that.''
\end{quote}

He concluded the testing session by stating:

\begin{quote}
``I am having fun, so I could do this actually quite a bit of time.''
\end{quote}

When asked for his second impressions, P3 expressed an overall positive evaluation of the system.
He emphasized the value of its structure, noting that even without AI-generated feedback, he would find it useful as a documentation tool due to its clarity and simplicity.
While acknowledging potential limitations over time, he stated that the system had been helpful during the session.

Regarding integration into his workflow, P3 indicated that he would primarily use the system in early design phases to concretize abstract, vibe-based ideas.
He described using the structure to clarify the purpose of individual elements before transferring the results into his existing documentation practices, with the possibility of returning to the system when introducing new features.

He concluded the interview by characterizing his stance toward the system:
\begin{quote}
    ``I am very positive towards the system, having my gripes with LLMs.
    I am highly critical, I am not a hater like many people are. 
    I think there is value, and I use it sometimes for different purposes. [...] And I think there is value for this problem domain here as well.''
\end{quote}

\paragraph{P4 Observations \& Second Impressions}
As P4 had already provided extensive feedback during the demonstration phase, he contributed fewer additional comments during this part of the interview.
P4 began by defining three design pillars before articulating the core design idea.
When receiving structural feedback on the final pillar, he chose to let the system generate a revised version, after which he stated:

\begin{quote}
``I don't feel like this version addresses this concern.''
\end{quote}

He subsequently prompted the system multiple times to identify and resolve structural issues.
After several iterations, he remarked:

\begin{quote}
``Now it's super watered down and abstract.''
\end{quote}

P4 then proceeded to the design evaluation functionality.
He noted that one identified contradiction could potentially be leveraged intentionally to create contrast (e.g., a cozy game featuring an evil bank), while also acknowledging that the feedback was generally useful.
Regarding suggested pillar additions, he commented:

\begin{quote}
``These are actually good suggestions, but for one [of them] it should recognize it's already there [as a pillar] and merge it.''
\end{quote}

He concluded the testing session by evaluating feedback on an intentionally contradictory feature, responding:

\begin{quote}
``Oh, it didn't say it's great, which is cool.''
\end{quote}

When asked for his second impressions, P4 described his perspective as largely unchanged from his initial reaction.
He reiterated concerns that the system frequently rephrased existing content without introducing substantially new perspectives, while noting its ability to surface contradictions and prompt reflection:

\begin{quote}
``It feels like it often just rephrases or rewords things that I already wrote somewhere [...] without genuinely finding new angles.
I do think it's decently good at finding contradictions with the things that it has there already.
I do think it makes you rethink what you put in there, just through the way of talking about it.''
\end{quote}

When asked whether the tool could fit into his professional workflow, P4 expressed reservations.
He stated that he currently preferred discussing design ideas with other people due to their contextual knowledge, while outlining conditions under which the tool could become more useful:

\begin{quote}
    ``If it was good at suggesting new and concrete things, I would really like it actually.
It could help with the idea by suggesting games that `do time manipulation like this' and `maybe we could do something like that'.''
\end{quote}

He concluded by reflecting on the broader context of the work, expressing appreciation for the academic exploration of such tools, while maintaining that current models were not yet sufficiently mature for practical game design use.

\paragraph{P5 Observations \& Second Impressions}
During the session, P5 provided minimal verbal commentary and did not consistently articulate her thoughts aloud, resulting in fewer observable verbal reactions.

P5 indicated that she would test the tool using an ongoing project for which she had not yet defined design pillars.
She began by describing the core design idea and subsequently created three pillars.
She then proceeded directly to the design evaluation functionality without first requesting structural feedback.

One suggested addition concerned a narrative pillar, which P5 had not previously defined.
She added a narrative pillar manually but did not adopt the suggested formulation.
She also tested the additional feature functionality using an intentional juxtaposition in the art style.
While the system generated feedback addressing this juxtaposition, P5 did not verbally comment on it and continued by testing the structural pillar feedback.

After several iterations of structural feedback and revisions, she remarked:

\begin{quote}
``Fixing it doesn't really change much. Every time I fix it, it's like adding more water.''
\end{quote}

When asked for her second impressions, P5 described the feedback on contradictions and additions positively, while expressing reservations about the volume and necessity of the information provided:

\begin{quote}
``The general feedback coverage and the contradictions and additions I think are pretty cool.
Though I don't think I need this much information if it's a good fit, because usually I already know what I am trying to do. [...]
But also the additional feature, that is actually quite nice, just to test against the pillars.''
\end{quote}

She further noted that the tool could be more useful for individuals not directly involved in defining the pillars, such as collaborators seeking to validate design decisions.

When asked whether she would use such a tool in practice, P5 responded conditionally.
She stated that she would use it to identify contradictions or potential additions, but would not rely on the structural pillar feedback, while reiterating her appreciation for the additional feature evaluation.

\paragraph{P6 Observations \& Second Impressions}
P6 began by entering the core design idea and immediately asked whether this element was supported by the system:

\begin{quote}
``I don't have any support on the [core design idea], right?
For another use case, I would like to have that.''
\end{quote}

He then added several design pillars and commented on his approach:

\begin{quote}
``Right now I am thinking the way I write those pillars is not perfect, but I am just gonna write it out and see what the LLM tells me.''
\end{quote}

After reviewing the AI-generated pillar feedback, P6 expressed concerns about how the system addressed perceived gaps:

\begin{quote}
``So kind of the problem I have with this is it just says `ok this is too broad, so let me just make up my own stuff to fill the gaps.'
Maybe it would make sense to include the design idea here, just so it doesn't come up with random stuff.''
\end{quote}

As he continued reading the feedback, P6 commented on the number of issues identified by the system, which led him to question how the feedback related to his use of design pillars.
When interacting with the design evaluation functionality, he stated:

\begin{quote}
``I have a problem with the contradictions.
Especially the second one to me seems more like a problem about the game idea itself... which is fair, but... kind of confusing. [...]
The things it says are valid points, but to me right now it feels more of a feedback on the game idea instead of the pillars, at least for the contradictions here.
I guess it kinda makes sense because the pillars describe the idea, right.''
\end{quote}

P6 subsequently tested the additional feature functionality with several ideas and responded positively to the resulting feedback:

\begin{quote}
``I would say the feature feedback is the best thing.
Like, everything it says makes sense to me and I agree with it.
This is definitely a thing that I could see myself using, especially because it's a more objective look on things.''
\end{quote}

When asked for his second impressions, P6 described the tool as potentially useful, while emphasizing the need for critical engagement with AI-generated feedback:

\begin{quote}
``Overall, I think it's a tool that can definitely be useful.
Like with everything that AI and LLMs give you, you have to take it with a grain of salt, but if you do that, [...] then definitely it can help you think about things you haven't thought about.
I can definitely see myself using this.''
\end{quote}

\paragraph{Feature Suggestions}
We concluded the interviews by asking whether there were any features the participants would have liked to see. 
The answers are presented in \autoref{tab:feature_requests}.
The table additionally includes (indirect) feature suggestions that came up during the demonstration or testing phase of the interview.

\section{Discussion}
We conducted two user-centered evaluations of SPINE.
First, we deployed the tool in an uncontrolled game jam setting, documented usage, and collected participants' first impressions.
Second, we evaluated the tool in a controlled setting through expert interviews.
Across both studies, the findings provide initial evidence that the tool’s workflow is feasible in practice and that practitioners see potential value in it, while also highlighting clear limitations in output quality and fit to individual workflows.
We discuss each evaluation in turn and conclude with implications that connect the results.

\paragraph{Case Study}
In the game jam case study, we addressed our questions from \autoref{sec:approach} based on our documentation and the participants' reflections.
Regarding pillar definition, both participants reported that the tool was practical and helpful for articulating design intentions, and that it supported ideation by helping them externalize thoughts and explore alternatives.
These observations suggest that LLM-supported mixed-initiative workflows can be promising in early-stage design activities.

For the decision-making functionality, the evidence is less conclusive.
One participant reported that the decision-making step worked well and provided suggestions for improvement.
However, the functionality was used only rarely during the jam, resulting in limited data for this aspect.
This may partially reflect the constrained scope and timeframe of a game jam project, which can reduce opportunities for revisiting and negotiating design decisions.
Based on our prior experiences, game jam development often involves frequent ideation and changing requirements.
In this case, the team appeared to align relatively early.
A plausible explanation is that defining pillars early, supported by the tool, contributed to a shared vision, and that the small team size and clear roles reduced coordination overhead.
Overall, even under these constraints, participants indicated that the functionality was at least somewhat helpful, but stronger evidence would require broader deployment.

The game jam study is also limited by its small scale and number of participants.
Nevertheless, we consider this approach appropriate at an early stage of investigating such systems, where rapid iteration and small-scale testing can help identify promising directions prior to more extensive evaluation.

\begin{table}[t!]
    \centering
    \caption{Feature Request Collection}
    \label{tab:feature_requests}
    \resizebox{\columnwidth}{!}{
    \begin{tabular}{p{0.7\columnwidth}c}
        \hline
        \textbf{Feature} & \textbf{Participant} \\
        \hline

        Might be interesting to train the model on games stuff & P3, P4 \\

        Nudge and give hints how to improve pillars instead of doing it for the user
        & P3, P6 \\

        Some form of documentation: What is a pillar? What is a core design idea? ...
        & P3, P5, P6 \\
        
        Button to add pillar from additions & P3, P6 \\

        Short reasoning for structural issues & P4, P5, P6 \\

        Pillar history to be able to better iterate and try out things & P4 \\

        Feedback on the content of the pillar in addition to the structure & P5 \\

        Contextual knowledge: Knowing about games in general, similar games with similar pillars, suggesting games that do something similar, compare idea to those games, ... & P4 \\

        Include core design idea in pillar rewriting process for higher quality & P6 \\

        General quality improvements of response, including more unique scenarios like juxtapositions & P3, P4, P5, P6 \\

        \hline
    \end{tabular}
    }
\end{table}

\paragraph{Expert Interviews}
The interviews provided deeper insights into how practitioners perceive the tool and its functionalities.
Overall reception was mixed, but participants consistently expressed interest in the underlying vision and described specific aspects as useful.
One participant responded very positively, two participants emphasized potential, and one participant was more critical, noting a general bias against LLMs and ethical concerns.
While this stance does not invalidate the findings, it highlights that adoption is not solely a question of functionality and may depend on preferences, values, and professional practices.
Notably, even the most critical participant described the tool as ``pretty decent'' for certain use cases.

Participants differed in their assessments of individual SPINE functionalities.
This variation may reflect differences in working styles and expectations, and it may also indicate that the system’s outputs are not consistently aligned with user intent.
Across interviews, the additional-feature evaluation was frequently described as particularly valuable, despite being among the least developed features.
This points to a promising direction for supporting reflective decision-making during design, and it also provides converging support for the more limited evidence from the game jam setting.

Limitations of the interview study include the moderate sample size and the focus on exploring functionality rather than measuring performance quantitatively.
We consider a qualitative approach warranted at this stage, as it supports in-depth analysis of perceived value, breakdowns, and opportunities for redesign.
These insights can inform future iterations of the system and motivate more rigorous evaluation designs.

\paragraph{System Limitations and Implications for Redesign}
A recurring limitation across both studies was perceived variability in output quality, which aligns with our analysis in \autoref{sec:model_performance}.
Participants also raised concerns about trust and transparency.
For example, one developer suggested that the system should be more transparent to increase trust, consistent with prior work~\cite{eigner2024determinants}.
Future iterations could address these issues by improving model selection and prompting, and by exploring approaches that better ground outputs in context (e.g., retrieval-augmented generation or agentic workflows).
Depending on feasibility and scope, domain adaptation (e.g., fine-tuning) may also be considered.

Finally, several participants noted that contradictions between pillars are not always undesirable and may sometimes be an intentional design choice that creates productive tension.
In the current system, mutual exclusivity assumptions led to feedback that did not reliably recognize such intentional juxtaposition.
A practical redesign implication is to soften contradiction feedback by explicitly accommodating intentional rule-breaking (e.g., flagging a contradiction as a risk while also inviting the designer to confirm whether it is deliberate and to articulate its purpose).

\paragraph{Design Pillars as Documentation Artifacts}
In addition to its mixed-initiative functionality, several participants noted that the tool was already valuable as a documentation aid.
They emphasized that the structured formulation of ideas and pillars was helpful in itself, independent of AI-generated feedback.
Formalized design pillars therefore hold potential beyond immediate decision support, enabling more standardized game design documentation, supporting clearer communication within teams, and laying the groundwork for more comprehensive, player--experience-driven design methodologies (e.g.,~\cite{dyrda2025toward}).
From this perspective, an AI-assisted documentation process centered on design pillars represents a promising step toward a unifying framework for pillar-based reasoning and decision-making in game development.

\section{Related Work}
An overview of the related work regarding game design pillars can be found in \autoref{sec:pillars}. This section focuses on the additional related work topics of this paper.

Decision support systems leveraging LLMs are an ongoing research topic in various domains, including medicine~\cite{park2025synergistic}, city management~\cite{kalyuzhnaya2025llm}, and entrepreneurship~\cite{doshi2025generative,csaszar2024artificial,alkayyal2025llm}.
Lubos et al.~\cite{lubos2025towards} built an LLM-based system that analyzes recorded group discussions to facilitate the decision-making process in teams and provide informed recommendations, which proved to extend basic meeting documentation with deeper insights.
Kalyuzhnaya et al.~\cite{kalyuzhnaya2025llm} researched how a multi-agent system could improve existing urban information systems and demonstrated the practical applicability of their approach.
While SPINE shares a common motivation with all these decision-making tools, none of the existing literature focuses on the domain of game design, which we address specifically through game design pillars.

In terms of general game design processes, the use of LLMs has received increasing attention~\cite{gallotta2024large,sweetser2024large}.
Lee et al.~\cite{lee2023empowering} explore AI-based game design workflows for generating complete game design proposals, including concept art and documentation.
Begemann et al.~\cite{begemann2024empirical} and Long et al.~\cite{long2024sketchar} investigate generative AI tools during early stages of game development, showcasing their potential to support creativity and concept generation.
In contrast, our study focuses on natural language artifacts from the early stages of game development.
Geheeb et al.~\cite{geheeb2025diamonds} built the toolkit \emph{SPARC}, which leverages medium-sized LLMs that can be run locally to refine game concepts based on 10 pre-defined aspects.
Their results indicate a positive reception among students, with some being eager to use such tools in the future.
We share the focus on natural language artifacts in early development.
As concept creation and pillar creation often go hand in hand, we even share the exact step in the process.
Nevertheless, our focus on game design pillars sets us apart and effectively complements the work of Geheeb et al., as it would allow designers to go back and forth between the concept and the pillars during the creation process of both.

\section{Conclusion and Future Work}
In this paper, we investigated the potential of large language models (LLMs) to support mixed-initiative game design pillar workflows.
We contributed a formal definition of game design pillars and presented SPINE, a mixed-initiative prototype designed to support both the creation and usage of pillars.
Through a small-scale case study conducted during a game jam and a qualitative interview study with four professional game developers, we explored how such a system is perceived and used in practice.
Across both evaluations, participants described LLM-based aspects of the tool as helpful, particularly for articulating design intent, reflecting on contradictions, and evaluating features against established pillars.
At the same time, they identified clear limitations related to output quality, consistency, and transparency.
Beyond its role in mixed-initiative decision support, our findings also point toward the value of design pillars as structured documentation artifacts.
Taken together, these results support our original framing: LLMs constitute a promising and appropriate candidate for mixed-initiative support in this domain, meriting further research and iterative refinement.
The observed benefits, breakdowns, and tensions illustrate both the opportunities and challenges of integrating LLMs into creative design practices, rather than resolving them.

Future work spans technical, interactional, and evaluative directions.
On the modeling side, output quality could be improved through prompt refinement, fine-tuning with annotated datasets, or the integration of additional contextual knowledge via retrieval-augmented generation (RAG) or agentic approaches.
From an interaction design perspective, several participants emphasized the value of ``nudging'' feedback that prompts reflection without overriding creative agency.
This aligns with prior guidance on pillar development as a process of structured questioning rather than prescription~\cite{despain2013100principles}, and represents a particularly promising direction for future iterations.
Additional work could explore how SPINE complements other LLM-based approaches for early-stage game design, such as concept development~\cite{geheeb2025diamonds}, as well as how participant-suggested features (see \autoref{tab:feature_requests}) affect usability and adoption.
Finally, future studies should evaluate more mature versions of the system with larger and more diverse participant groups to better understand how such tools integrate into different professional workflows.
Overall, this work provides an initial step toward understanding how LLM-supported, mixed-initiative systems can assist game designers in reasoning about design pillars in an early stage of research.


\bibliographystyle{ACM-Reference-Format}
\bibliography{bibliography}

@inproceedings{zagal2023considering,
  title={Considering Large Student Teams in Game Development Education: A Post-Mortem},
  author={Zagal, Jose},
  booktitle={Conference Proceedings of DiGRA 2023 Conference: Limits and Margins of Games Settings},
  year={2023}
}

@inproceedings{luo2021multidisciplinary,
  title={A multidisciplinary approach To designing immersive gameplay elements for learning standard-based educational content},
  author={Luo, Vinson and Klinkert, Lawrence J and Foster, Paul and Tseng, Ching-Yu and Adams, Elizabeth and Ketterlin-Geller, Leanne and Larson, Eric C and Clark, Corey},
  booktitle={Extended Abstracts of the 2021 Annual Symposium on Computer-Human Interaction in Play},
  pages={67--73},
  year={2021}
}

@book{despain2013100principles,
  title        = {100 Principles of Game Design},
  editor       = {Despain, Wendy},
  contributors = {Acosta, Keyvan and Canacari-Rose, Liz and Deneen, Michael and Hiwiller, Zach and Howard, Jeff and Kadinger, Christina and Keeling, Chris and Kuczik, Casey},
  publisher    = {New Riders (an imprint of Peachpit, a division of Pearson Education)},
  address      = {Berkeley, CA},
  year         = {2013},
  isbn         = {978-0-321-90249-8}
}

@inproceedings{lai2020towards,
  title={Towards friendly mixed initiative procedural content generation: Three pillars of industry},
  author={Lai, Gorm and Latham, William and Leymarie, Frederic Fol},
  booktitle={Proceedings of the 15th International Conference on the Foundations of Digital Games},
  pages={1--4},
  year={2020}
}

@misc{keenan2017duskers,
  author       = {Keenan, Tim},
  title        = {Finding Duskers: Innovation Through Better Design Pillars},
  howpublished = {Video recording, Game Developers Conference 2017},
  month        = {March},
  year         = {2017},
  note         = {YouTube video, accessed 27 October 2025, https://www.youtube.com/watch?v=kzQDVtysXjA}
}

@misc{cleveland2019subnautica,
  author       = {Cleveland, Charlie},
  title        = {The Design of \textit{Subnautica}},
  howpublished = {Video recording, Game Developers Conference 2019},
  month        = {March},
  year         = {2019},
  note         = {YouTube video, accessed 27 October 2025, https://www.youtube.com/watch?v=7R-x9NSBS2Y}
}

@misc{davis2018godofwar,
  author       = {Davis, Rob},
  title        = {The Level Design of \emph{God of War}},
  howpublished = {Video recording, Game Developers Conference 2018},
  month        = {March},
  year         = {2018},
  note         = {YouTube video, accessed 27 October 2025, https://www.youtube.com/watch?v=eSB29qx6sWw}
}

@misc{ParadoxInteractive_GamePillars,
  author       = {Paradox Interactive},
  title        = {Game Pillars – What makes a game a Paradox game},
  howpublished = {Web page},
  year         = {n.d.},
  note         = {\url{https://www.paradoxinteractive.com/our-company/our-business/game-pillars} (accessed 27 October 2025)}
}

@article{graft2012diablo,
  author       = {Kris Graft},
  title        = {The Devil’s Workshop: An Interview with Diablo III’s Jay Wilson},
  journal      = {Game Developer},
  year         = {2012},
  month        = {May},
  day          = {14},
  url          = {https://www.gamedeveloper.com/design/the-devil-s-workshop-an-interview-with-i-diablo-iii-i-s-jay-wilson},
  note         = {Accessed: 2025-11-05}
}

@misc{lapikas2012deusex,
  author       = {Lapikas, Fran\c{c}ois},
  title        = {Reimagining a Classic: The Design Challenges of \emph{Deus Ex: Human Revolution}},
  howpublished = {Video recording, Game Developers Conference 2012},
  month        = {February},
  year         = {2017},
  note         = {YouTube video, accessed 05 November 2025, https://www.youtube.com/watch?v=I5wwviUJV9M}
}

@misc{ali2013destiny,
  author       = {Ali, Orry},
  title        = {Destiny: Bungie’s Brave New Worlds — An In-Depth Look at \emph{Destiny}},
  howpublished = {Online article, Polygon},
  month        = {February},
  day          = {17},
  year         = {2013},
  note         = {Accessed 05 November 2025, https://www.polygon.com/2013/2/17/3993058/destiny-bungie-first-look-preview/}
}

@misc{cain2023fallout,
  author       = {Timothy Cain},
  title        = {Design Pillars},
  howpublished = {Video recording, YouTube},
  year         = {2023},
  note         = {YouTube video, accessed 05 November 2025, https://www.youtube.com/watch?v=N7b7LFXBZ9M}
}

@misc{Pears2017,
  author       = {Pears, Max},
  title        = {Design Pillars – The Core of Your Game},
  howpublished = {Web page},
  month        = {September},
  day          = {2},
  year         = {2017},
  note         = {\url{https://www.maxpears.com/2017/09/02/design-pillars-the-core-of-your-game/} (accessed 27 Oct 2025)}
}

@misc{Wagar_DesignPillars_GameDesignSkills,
  author       = {Wagar, Celia},
  title        = {Game Design Pillars: What Are They and How to Practically Apply Them},
  howpublished = {Web page},
  year         = {2023},
  note         = {\url{https://gamedesignskills.com/game-design/design-pillars/} (accessed 27 October 2025)}
}

@article{geheeb2025diamonds,
  title={Diamonds in the rough: Transforming SPARCs of imagination into a game concept by leveraging medium sized LLMs},
  author={Geheeb, Julian and Ivan, Farhan Abid and Dyrda, Daniel and Ansch{\"u}tz, Miriam and Groh, Georg},
  booktitle    = {Proceedings of AI4HGI ’25: The First Workshop on Artificial Intelligence for Human-Game Interaction at the 28th European Conference on Artificial Intelligence (ECAI ’25)},
  address      = {Bologna, Italy},
  month        = {October},
  year         = {2025},
}

@online{kara2021GamePillars,
  author       = {Kara},
  title        = {Game Pillars: Set Limits To Your Game Direction To Focus Your Design},
  year         = {2021},
  url          = {https://www.whalebraindesign.com/newsletter/what-are-game-pillars},
  note         = {Accessed: 2025-11-04}
}

@inproceedings{ledo2018evaluation,
  title={Evaluation strategies for HCI toolkit research},
  author={Ledo, David and Houben, Steven and Vermeulen, Jo and Marquardt, Nicolai and Oehlberg, Lora and Greenberg, Saul},
  booktitle={Proceedings of the 2018 CHI conference on human factors in computing systems},
  pages={1--17},
  year={2018}
}

@online{ahmad2025BasicPillarsSystemDesign,
  author       = {Ahmad, Arslan},
  title        = {4 Basic Pillars of System Design – Scalability, Availability, Reliability, Performance},
  year         = {2025},
  url          = {https://www.designgurus.io/blog/4-basic-pillars-of-system-design},
  note         = {Accessed: 2025-11-20}
}

@online{idunnu_paul2024FantasticFourSystemDesign,
  author       = {Idunnu Paul, Joshua},
  title        = {The Fantastic Four of System Design: Scalability, Availability, Reliability and Performance},
  year         = {2024},
  url          = {https://cybernerdie.medium.com/the-fantastic-four-of-system-design-scalability-availability-reliability-and-performance-ef247cd4bd2c},
  note         = {Accessed: 2025-11-20}
}

@article{eigner2024determinants,
  title={Determinants of llm-assisted decision-making},
  author={Eigner, Eva and H{\"a}ndler, Thorsten},
  journal={arXiv preprint arXiv:2402.17385},
  year={2024}
}

@article{park2025synergistic,
  title={Synergistic joint model of knowledge graph and llm for enhancing xai-based clinical decision support systems},
  author={Park, Chaelim and Lee, Hayoung and Lee, Seonghee and Jeong, Okran},
  journal={Mathematics},
  volume={13},
  number={6},
  pages={949},
  year={2025},
  publisher={MDPI}
}

@inproceedings{lubos2025towards,
  title={Towards Group Decision Support with LLM-based Meeting Analysis},
  author={Lubos, Sebastian and Felfernig, Alexander and Garber, Damian and Le, Viet-Man and Henrich, Manuel and Willfort, Reinhard and Fuchs, Jeremias},
  booktitle={Adjunct Proceedings of the 33rd ACM Conference on User Modeling, Adaptation and Personalization},
  pages={331--335},
  year={2025}
}

@article{doshi2025generative,
  title={Generative artificial intelligence and evaluating strategic decisions},
  author={Doshi, Anil R and Bell, J Jason and Mirzayev, Emil and Vanneste, Bart S},
  journal={Strategic Management Journal},
  volume={46},
  number={3},
  pages={583--610},
  year={2025},
  publisher={Wiley Online Library}
}

@article{csaszar2024artificial,
  title={Artificial intelligence and strategic decision-making: Evidence from entrepreneurs and investors},
  author={Csaszar, Felipe A and Ketkar, Harsh and Kim, Hyunjin},
  journal={Strategy Science},
  volume={9},
  number={4},
  pages={322--345},
  year={2024},
  publisher={INFORMS}
}

@article{lee2023empowering,
  title={Empowering game designers with generative AI},
  author={Lee, J and Eom, So-Youn and Lee, J},
  journal={IADIS International Journal on Computer Science \& Information Systems},
  volume={18},
  number={2},
  pages={213--230},
  year={2023}
}

@article{gallotta2024large,
  title={Large language models and games: A survey and roadmap},
  author={Gallotta, Roberto and Todd, Graham and Zammit, Marvin and Earle, Sam and Liapis, Antonios and Togelius, Julian and Yannakakis, Georgios N},
  journal={IEEE Transactions on Games},
  year={2024},
  publisher={IEEE}
}

@inproceedings{sweetser2024large,
  title={Large language models and video games: A preliminary scoping review},
  author={Sweetser, Penny},
  booktitle={Proceedings of the 6th ACM Conference on Conversational User Interfaces},
  pages={1--8},
  year={2024}
}

@article{long2024sketchar,
  title={Sketchar: supporting character design and illustration prototyping using generative AI},
  author={Long, LING and Xinyi, CHEN and Ruoyu, WEN and Toby Jia-Jun, LI and Ray, LC},
  journal={Proceedings of the ACM on Human-Computer Interaction},
  volume={8},
  number={CHI PLAY},
  pages={337},
  year={2024}
}

@article{begemann2024empirical,
  title={Empirical insights into AI-assisted game development: A case study on the integration of generative AI tools in creative pipelines},
  author={Begemann, Andrew and Hutson, James},
  journal={Metaverse},
  volume={5},
  number={2},
  year={2024}
}

@article{kalyuzhnaya2025llm,
  title={LLM Agents for Smart City Management: Enhancing Decision Support Through Multi-Agent AI Systems.},
  author={Kalyuzhnaya, Anna and Mityagin, Sergey and Lutsenko, Elizaveta and Getmanov, Andrey and Aksenkin, Yaroslav and Fatkhiev, Kamil and Fedorin, Kirill and Nikitin, Nikolay O and Chichkova, Natalia and Vorona, Vladimir and others},
  journal={Smart Cities (2624-6511)},
  volume={8},
  number={1},
  year={2025}
}

@inproceedings{dyrda2025toward,
  title={Toward a Game Design Engineering Process Centered on Player Experience},
  author={Dyrda, Daniel and Klinker, Gudrun},
  booktitle={2025 IEEE Conference on Games (CoG)},
  pages={1--4},
  year={2025},
  organization={IEEE}
}

@inproceedings{alkayyal2025llm,
  title={An LLM-Based Decision Support System for Strategic Decision-Making},
  author={Alkayyal, Majd and Malberg, Simon and Groh, Georg},
  booktitle={Joint European Conference on Machine Learning and Knowledge Discovery in Databases},
  pages={460--464},
  year={2025},
  organization={Springer}
}

@inproceedings{Dyrda2026GameDesignPillars,
  author    = {Dyrda, Daniel and Wink Rodrigues Lucas, Felipe and Schacherbauer, Martin and Bika, Chrysa and Geheeb, Julian and Pirker, Johanna},
  title     = {Game Design Pillars: Between Concept and Practice},
  booktitle = {Proceedings of the Foundations of Digital Games Conference (FDG '26)},
  year      = {2026},
  address   = {Copenhagen, Denmark},
  publisher = {Association for Computing Machinery},
  doi       = {10.1145/3815598.3815686},
  isbn      = {979-8-4007-2495-4/2026/08}
}

\appendix
\onecolumn
\section{Game Design Pillar Dataset}\label{sec:pillar_set}
\begin{table*}[h]
  \caption{Game: Subnautica, Credibility: high~\cite{cleveland2019subnautica} (Timestamp 10:00)}
  \label{tab:subnautica}
  \begin{tabular}{p{0.25\textwidth}p{0.7\textwidth}}
    \toprule
    Title & Description\\
    \midrule
    Vessel Design/Building\newline(Intoxicating Creation) & The overwhelming excitement of being able to build anything. "I do not think there is any thrill that can go through the human heart like that felt by the inventor as he sees some creation of the brain unfolding to success... such emotions make a man forget food, sleep, friends, love, everything." - Nikola Tesla (1943). Barn raising?\\
    Exploration, Discovery\newline(Thrill of the Unknown) & Excitement, dread and tension of exploring the unknown. No idea what dangers/rewards are down there. Increased risk generally associated with increased reward.\\
    Challenge, Teamwork, Interdependent Systems\newline(Cascading Hysteria) & (uncontrollable outburst of emotion, fear, irrationality, laughter, weeping, etc.) - FTL style chain of "oh shit" dependencies, where a failure in one system can affect others, until you're suddenly in trouble (sensors stops working, can't see enemies onboard). Creates urgency, drama, impetus, teamwork, interrelatedness.\\
  \bottomrule
  \end{tabular}
\end{table*}

\begin{table*}[h]
  \caption{Game: God of War, Credibility: high~\cite{davis2018godofwar} (Timestamp (5:00))}
  \label{tab:godofwar}
  \begin{tabular}{p{0.25\textwidth}p{0.7\textwidth}}
    \toprule
    Title & Description\\
    \midrule
    Combat & \begin{itemize}
        \item Marvel Films vs Marvel Comics
        \item New Mythical Creatures/Weapons
        \item New Midrange \& Close Combat
        \item High Optical Mocap Fidelity
        \item Integrated Son Companion
    \end{itemize}\\
    Father \& Son\newline (Narrative) & \begin{itemize}
        \item Humanize Kratos
        \item Father teaches son; Son teaches Father
        \item Son helps not hinder Kratos
        \item Son is Believable
        \item Symbiotic Relationship
    \end{itemize}\\
    Exploration & \begin{itemize}
        \item Discovery
        \item Resource gathering
        \item Hunt and Make
        \item Father and Son Bonding
        \item Cerebral Engagement
    \end{itemize}\\
  \bottomrule
  \end{tabular}
\end{table*}


\begin{table*}[h]
  \caption{Game: Dusker's, Credibility: high~\cite{keenan2017duskers} (Timestamp (5:00))}
  \label{tab:duskers}
  \begin{tabular}{p{0.25\textwidth}p{0.7\textwidth}}
    \toprule
    Title & Description\\
    \midrule
    Realism & - (Not the actual text: The game feels tight, the game feels real)\\
    Isolation & - (Not the actual text: You are alone)\\
    Planning & - \\
  \bottomrule
  \end{tabular}
\end{table*}



\begin{table*}[h]
  \caption{Diablo III, Credibility: high~\cite{graft2012diablo}}
  \label{tab:diablo}
  \begin{tabular}{p{0.45\textwidth}p{0.5\textwidth}}
    \toprule
    Title & Description\\
    \midrule
    Approachable & - \\
    Powerful heroes & - \\
    Highly customizable & - \\
    Great item game & - \\
    Endlessly replayable & - \\
    Strong setting & - \\
    Cooperative multiplayer & - \\
  \bottomrule
  \end{tabular}
\end{table*}

\begin{table*}[h]
  \caption{Deus Ex: Human Revolution, Credibility: high~\cite{lapikas2012deusex} (Timestamp 15:30)}
  \label{tab:deusex}
  \begin{tabular}{p{0.25\textwidth}p{0.7\textwidth}}
    \toprule
    Title & Description\\
    \midrule
    It's about choice & This can be explained best by the "What if?" scenario. As the player explores the game and comes up against challenges, he'll be asking himself "what if I do this?" or "what if I try that?" And each time the game should answer back: "Yes, that is possible." \\
    Every choice has a consequence & Every time the player makes a choice, there should be a set of consequences associated with it, both good and bad. \\
    Make it simpler & Now, our goal is not to dumb it down. We have no illusions, a Deus Ex game will never be as simple to play as a Halo. But by streamlining some features, we are confident we can create a game that is both elegantly manageable and right a[t the same tim]e. Metroid Prime is a good example of this type of design philosophy. \\
    Make it spectacular and rewarding & So our goal is not only to make a game that is fun and open, we also want the player's actions to have a visceral and gratifying feeling to them. He shouldn't just do something [because it is] useful, he should also do it becaues the \textit{likes} it. \\
    \midrule
    Combat & - \\
    Stealth & - \\
    Hacking & - \\
    Social & - \\
  \bottomrule
  \end{tabular}
\end{table*}

\begin{table*}[h]
  \caption{Destiny, Credibility: high~\cite{ali2013destiny}}
  \label{tab:destiny}
  \begin{tabular}{p{0.45\textwidth}p{0.5\textwidth}}
    \toprule
    Title & Description\\
    \midrule
    A world players want to be in & - \\
    A bunch of fun things to do & - \\
    Rewards players care about & - \\
    A new experience every night & - \\
    Shared with other people & - \\
    Enjoyable by all skill levels & - \\
    Enjoyable by the tired, impatient and distracted & - \\
  \bottomrule
  \end{tabular}
\end{table*}

\begin{table*}[h]
  \caption{Fallout, Credibility: high~\cite{cain2023fallout} (Timestamp 5:50)}
  \label{tab:fallout}
  \begin{tabular}{p{0.45\textwidth}p{0.5\textwidth}}
    \toprule
    Title & Description\\
    \midrule
    Mega levels of violence & - \\
    There is often no right solution & - \\
    There should always be multiple solutions & - \\
    Player's actions affect the world, and the world will react to the player & - \\
    The game should be open-ended & - \\
    The player should always have a goal & - \\
    The player has control over his own actions & - \\
    Interface & - \\
    Encounter Windows & - \\
    Wide variety of weapons and armor and actions the player could take with them & - \\
    Detailed character creation rules, but also pre-made characters & - \\
    Make this game for the public, but make people who play GURPS happy & - \\
    The team is very motivated & - \\
  \bottomrule
  \end{tabular}
\end{table*}
\begin{table*}[h]
  \caption{Arcanum, Credibility: high~\cite{cain2023fallout} (Timestamp 12:15)}
  \label{tab:arcanum}
  \begin{tabular}{p{0.2\textwidth}p{0.7\textwidth}}
    \toprule
    Title & Description\\
    \midrule
    Tech that matters & - \\
    Rich, class-free character creation & - \\
    A complex, stat-driven game system & - \\
    A huge, single-player questline & - \\
    Multiplayer & - \\
  \bottomrule
  \end{tabular}
\end{table*}
\begin{table*}[h]
  \caption{The Outer Worlds, Credibility: high~\cite{cain2023fallout} (Timestamp 16:10)}
  \label{tab:outerworlds}
  \begin{tabular}{p{0.2\textwidth}p{0.7\textwidth}}
    \toprule
    Title & Description\\
    \midrule
    Simple, but deep & - \\
    Dark, but humorous & - \\
    Fun trumps realism, but be consistent & - \\
    Classic Obsidian Role-Playing & - \\
  \bottomrule
  \end{tabular}
\end{table*}

\begin{table*}[h!]
  \caption{Publisher: Paradox Interactive, Credibility: high~\cite{ParadoxInteractive_GamePillars}}
  \label{tab:dataset}
  \begin{tabular}{p{0.2\textwidth}p{0.7\textwidth}}
    \toprule
    Title & Description\\
    \midrule
     Agency & Paradox games give players the freedom to live out their fantasies, create their own stories and express themselves and their creativity. From customization options, game rules and modding to emergent stories and rewriting history, Paradox titles are not linear, plot driven experiences.\\
     Living Worlds & Paradox games feature dynamic, reactive worlds where other forces seem to be pursuing their own goals beyond the control of players. No two games will be the same and players will experience new stories every time they play.\\
     Inviting & Paradox games have compelling themes with a clear promise. Players are enticed to make the effort of learning our games. (Which are approachable enough to keep players engaged, onboarding them in their worlds and gameplay systems in a smooth and rewarding way.)\\
     Cerebral & Paradox games challenge the player’s mind before their reflexes. Incredible depth rewards the player’s curiosity and intelligence. Our games are hard to master; there is always more to discover. Moreover, players can “nerd out” on the themes and subject matters even when not playing.\\
     Endless Experiences & One does not simply “finish” a Paradox game. Either you keep coming back for another playthrough or there is no end state at all. Paradox games provide engagement for a long time.\\
  \bottomrule
  \end{tabular}
\end{table*}



\twocolumn

\section{Prompt Templates Used in the Study}\label{sec:prompts}

This appendix lists the exact prompt templates used for evaluating and refining Game Design Pillars. These prompts were provided to the language model without modification, except for runtime substitution of placeholder variables (e.g., game ideas, pillar names, and descriptions).

\subsection{Validation Prompt}

\begin{verbatim}
Validate the following Game Design Pillar.
Check for structural issues regarding the following 
points:
1. The name does not match the description.
2. The intent of the pillar is not clear.
3. The pillar focuses on more than one aspect.
4. The description uses bullet points or lists.
Name: %s
Description: %s
For each feedback limit your answer to one sentence.
Answer as if you were talking directly to the designer.
\end{verbatim}

\subsection{Pillar Improvement Prompt}

\begin{verbatim}
Improve the following Game Design Pillar.
Check for structural issues regarding the following 
points:
1. The title does not match the description.
2. The intent of the pillar is not clear.
3. The pillar focuses on more than one aspect.
4. The description uses bullet points or lists.
Pillar Title: %s
Pillar Description: %s
Rewrite erroneous parts of the pillar and return a new 
pillar object.
\end{verbatim}

\subsection{Pillar Completeness Prompt}

\begin{verbatim}
Assume the role of a game design expert.
Evaluate if the following Game Design Pillars are a good 
fit for the game idea, explain why.
Also check if the pillar contradicts the direction of the 
game idea.

Game Design Idea: %s

Design Pillars: %s
\end{verbatim}

\subsection{Pillar Contradiction Prompt}

\begin{verbatim}
Assume the role of a game design expert.
Evaluate if the following Game Design Pillars stand in 
contradiction towards each other. Use the Game Design Idea
as context.

Game Design Idea: %s

Design Pillars: %s
\end{verbatim}

\subsection{Pillar Addition Prompt}

\begin{verbatim}
Assume the role of a game design expert.
Evaluate if the following Game Design Idea is sufficiently
covered by the following Game Design Pillars.

Game Design Idea: %s

Design Pillars: %s
If not, add new pillars to cover the missing aspects.
\end{verbatim}

\subsection{Context Alignment Prompt}

\begin{verbatim}
Assume the role of a game design expert.
Evaluate how well the following idea aligns with the 
given Game Design Pillars.

Idea: %s

Design Pillars: %s
\end{verbatim}

\end{document}